\documentstyle[sprocl,psfig]{article}

\def\Journal#1#2#3#4{{#1} {\bf #2}, #3 (#4)}

\def\NPB{{\em Nucl. Phys.} B}

\def\PLB{{\em Phys. Lett.} B}

\def\PRC{{\em Phys. Rev.} C}

\begin{document}

\title{\hfill TRI-PP-98-28\\[0.2cm]
EFFECTIVE THEORY FOR THE NON-RELATIVISTIC THREE-BODY SYSTEM
}

\author{H.-W. HAMMER}
\address{TRIUMF, 4004 Wesbrook Mall, Vancouver, B.C., Canada V6T 2A3 }


\maketitle

\abstracts{
We discuss renormalization of the
non-relativistic three-body problem with short-range forces.
The problem becomes non-perturbative at momenta of the
order of the inverse of the
two-body scattering length, and an infinite number of
graphs must be summed.
This summation leads to a cutoff dependence
that does not appear in any order in perturbation theory. We argue
that this cutoff dependence can be absorbed in a single three-body 
counterterm and compute
the running of the three-body force with the cutoff.}

There has been considerable interest recently in applying the successful 
concept of effective field theory (EFT) to nuclear physics\cite{edict}.
If the momenta $k$ are small compared to the inverse range of the 
interaction $1/R$, EFT provides a systematic expansion in powers of $k R$.
More complicated are systems where an unnaturally large parameter appears.
Specifically, for systems made out of nucleons and of $^4$He atoms,
the two-body scattering length $a_2$ is much larger than $R$.
In this case the expansion becomes
non-perturbative at momenta of the order of $1/a_2$,
in the sense that an infinite number of diagrams
must be resummed.
This resummation generates a new expansion in powers of $k R$ where
the full dependence in $k a_2$ is kept.
Consequently, the EFT is valid beyond $k\sim 1/a_2$, comprising,
in particular, bound states of size $\sim a_2$.
While there has been enormous progress in the two-body case\cite{edict}, 
the extension to three-particle systems presents us with a puzzle.
Although in some fermionic channels the resummed leading two-body
interactions lead to unambiguous
and very successful predictions\cite{23stooges},
amplitudes in bosonic systems and other fermionic channels
show sensitivity to the ultraviolet (UV) cutoff,
as is evidenced in the well known Thomas and Efimov effects.
This happens even though each leading-order three-body diagram with
resummed two-body interactions
is individually UV finite.
Below we  will argue that the EFT program can be extended to
three-boson systems with large $a_2$ by introducing a one-parameter 
three-body force counterterm at leading order\cite{letter}.

The most general Lagrangian involving a non-relativistic
boson $\psi$ and invariant under
small-velocity Lorentz, parity, and time-reversal transformations
can conveniently be written by introducing a dummy field $T$ with quantum
numbers of two bosons\cite{transvestite},
\begin{equation}
\label{lagt}
{\cal L}  =  \psi^\dagger(i\partial_{0}+\frac{\vec{\nabla}^{2}}{2M})\psi
         + \Delta T^\dagger T
         -\frac{g}{\sqrt{2}} (T^\dagger \psi\psi +\mbox{h.c.})
         +h T^\dagger T \psi^\dagger\psi +\ldots\;.
\end{equation}

\begin{figure}[ht]
\centerline{\psfig{figure=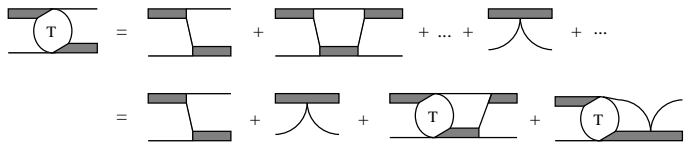,width=9cm}}
\caption{Diagrams contributing to particle/bound-state scattering.}
\label{fig1}
\end{figure}

We consider particle/bound-state scattering.
The diagrams contributing to this process in leading order
are illustrated in Fig. \ref{fig1}.
All diagrams including only two-body interactions are of the same
order for $k\sim 1/a_2$ and have to be summed.
This can be accomplished by solving the equation
represented by the second equality in Fig. \ref{fig1}\cite{23stooges,skorny}:
\begin{equation}
\label{aeq}
a(p)
=K(p,k)+\frac{2}{\pi}\int_0^\Lambda dq\ K(p,q)
\frac{q^2}{q^2-k^2-i\epsilon} a(q),
\end{equation}
\noindent
with $k$ ($p$) the incoming (outgoing) momentum,
$M E = 3k^2/4 - 1/a_2^2$ the total energy, 
$a(p=k)$ the scattering amplitude normalized
in such a way that $a_3=-a(0)$ is the particle/bound-state scattering
length, and
\begin{equation}
\label{kern}
K(p,q)=\frac{4}{3}(\frac{1}{a_2}+\sqrt{\frac{3}{4}p^2-M E})
    \left[\frac{1}{pq}{\rm ln}
    \left(\frac{q^2+p q+p^2-M E}
               {q^2-q p+p^2-M E}\right)
    +\frac{h}{Mg^2} \right].
\end{equation}
\noindent
The parametric dependence of $a(p)$ on $k$ is kept implicit. 
Three nucleons in the spin $J=1/2$ channel obey a pair
of integral equations with similar properties.
For $h=0$ and $\Lambda \to \infty$, the asymptotic solution of Eq. 
(\ref{aeq}) can be obtained analytically\cite{danilov}.
It turns out that the phase of the solution is
undetermined. For a finite $\Lambda$, however,
the solution has a well determined phase
which in the intermediate region $1/a_2\ll p\ll \Lambda$
is,
\begin{equation}
a(p)=A \cos(s_0 \ln \frac{p}{\Lambda} + \delta), \label{asym}
\end{equation}
\noindent
with $\delta$ is some dimensionless, cutoff-independent number.
Obviously, the limit $\Lambda\rightarrow\infty$ is not well defined.
Numerical solutions of
Eq. (\ref{aeq}) with $k=0$ for different values of $\Lambda$
confirm that the behavior
of $a(p)$ in the region $1/a_2\ll p \ll \Lambda$ is given by Eq.
(\ref{asym}) and that small differences in
the asymptotic phase lead to large differences in the particle/bound-state
scattering length\cite{letter}.

This cutoff dependence comes from the amplitude in the 
UV region, where the EFT Lagrangian, Eq. (\ref{lagt}), is not
to be trusted. In an EFT, the cutoff-dependent
contributions from high loop momenta are cancelled by counterterms in the 
Lagrangian and all uncertainty from the UV behavior of the theory is
parametrized by a few local counterterms.
Writing $h(\Lambda)=2M g^2 H(\Lambda)/\Lambda^2$ and assuming 
$H(\Lambda)\sim 1$, it is straightforward to see that the
term proportional to $H$ in Eq. (\ref{aeq}) becomes important 
only for $p\sim \Lambda$.
The asymptotic form, Eq.
(\ref{asym}), is still correct in the intermediate region.
A finite value of $H$ only changes the values of the amplitude
$A$ and the phase $\delta$, which become
functions of $H$, as is confirmed numerically\cite{letter}.
If $H$ is chosen to be a function of $\Lambda$
such as to cancel the explicit $\Lambda$ dependence,
we can make the solution of Eq. (\ref{aeq})
cutoff independent for all $p\ll \Lambda$.
In particular, the on-shell scattering amplitude 
$a(k)$ with $k\sim 1/a_2$ will be cutoff independent as well.
Thus $H(\Lambda)$ must be chosen such that
$-s_0 \ln \Lambda + \delta(H(\Lambda))= -s_0 \ln \Lambda_\star$,
where $\Lambda_\star$ is a parameter fixed by experiment or by matching
with a microscopic model.

\begin{figure}[ht]
\centerline{\psfig{figure=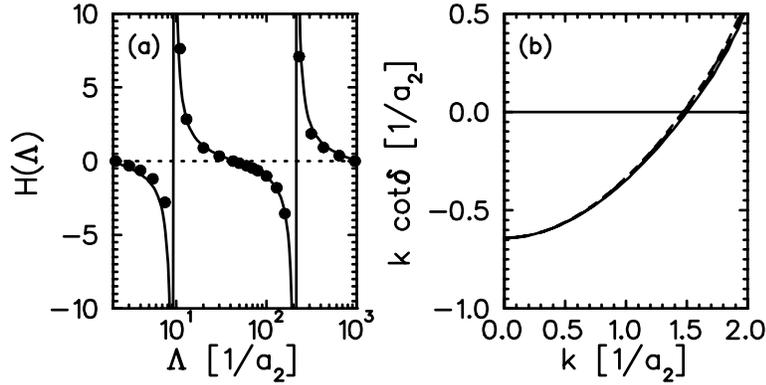,width=10cm}}
\caption{(a) Three-body force as function of $\Lambda$: numerical solution
(dots) and Eq. (\ref{h}). (b) Energy dependence: $k\cot\delta$ for different
cutoffs ($\Lambda=42.6,\,100.0,\,230.0,\,959.0\times a_2^{-1}$).}
\label{fig4}
\end{figure}

We can get a handle on the form of $H(\Lambda)$ by
considering Eq. (\ref{aeq}) with two different values of the cutoff
$\Lambda$ and $\Lambda '>\Lambda$.
Assuming both solutions have the same phase $\cos(s_0 \ln(p/\Lambda_\star))$
even for $p\sim\Lambda'$, we find\cite{letter}
\begin{equation}
\label{h}
H(\Lambda)= -\frac{\sin(s_0\ln({\Lambda}/{\Lambda_\star})-{\rm arctg}(1/s_0))}
                 {\sin(s_0 \ln({\Lambda}/{\Lambda_\star})+{\rm arctg}(1/s_0))}.
\end{equation}
\noindent
Consequently, with $H(\Lambda)$ chosen like Eq. (\ref{h}), 
the on-shell scattering amplitude
$a(k)$ for $k\ll\Lambda$ will be $\Lambda$ independent.
We also determine $H(\Lambda)$ numerically by finding the value
of $H$ that keeps the three-body scattering length $a_3=-a(0)$
constant for each value of $\Lambda$ varying over a large range.
The numerical values for $H(\Lambda)$ agree with Eq. (\ref{h}) to 
high accuracy (see Fig. \ref{fig4}(a)).
For illustration we used $a_3= 1.56 a_2$, but we have verified that similar
agreement holds for other values of $a_3$. In Fig. 
\ref{fig4}(b) we show the corresponding $k \cot\delta=i k +a(k)^{-1}$,
where $\delta$ is the $S$-wave phase shift for particle/bound-state
scattering, for several values of $\Lambda$.
As argued above, it is insensitive to
$\Lambda$ as long as $k\ll \Lambda$.
The effective range, e.g., is predicted as $r_3= 0.57 a_2$.
These arguments hold for the bound-state problem as well.
The shallowest bound state has a cutoff-independent
binding energy of $B_3=1.5/M a_2^2$.

The ratio $a_3/a_2=1.56$ is suggested by the values
$a_2=124.7$ \AA \ and $a_3=195$ \AA \
obtained from a phenomenological $^4$He-$^4$He potential\cite{heliumpot} 
giving the correct dimer binding energy.
Fig. \ref{fig4}(b) then represents the
phase shifts for atom/dimer scattering, with an effective
range $r_3= 71$ \AA.
Similarly, our result for the shallowest bound state
suggests an excited state of the trimer at $B_3=1.2$ mK.
Because of similar integral equations,
our arguments are relevant for three-fermion systems
with internal quantum numbers as well\cite{more3stooges}.

In conclusion,
we have provided evidence that
renormalization of the three-body problem with short-range forces
requires in general
the presence of a leading order one-parameter contact three-body force.

The work presented in this talk was done in collaboration with
P.F. Bedaque and U. van Kolck. This research was supported by
the Natural Science and Engineering Research Council of Canada.

\end{document}